\documentclass[twocolumn,showpacs,preprintnumbers,prl,amsmath,amssymb]{revtex4}

\usepackage{graphicx}
\usepackage{epsfig} 
\usepackage{dcolumn}
\usepackage{bm}

\begin{document}
\newcommand{\rth}{$R_{th}$ }
\newcommand{\rthy}{$R_{th}$}
\newcommand{\cll}{$c_{11}$ }
\newcommand{\clly}{$c_{11}$}
\newcommand{\cuu}{$c_{66}$ }
\newcommand{\cuuy}{$c_{66}$}
\newcommand{\cms}{cm$^{2}$ }
\newcommand{\cmsy}{cm$^{2}$}
\newcommand{\ab}{$\sim$ }
\newcommand{\aby}{$\sim$}
\newcommand{\tp}{$T'$ }
\newcommand{\tpy}{$T'$}
\newcommand{\tph}{$T_{p}$ }
\newcommand{\tphy}{$T_{p}$}
\newcommand{\too}{$T_{0}$ }
\newcommand{\tooy}{$T_{0}$}
\newcommand{\tauep}{$\tau_{ep}$ }
\newcommand{\tauepy}{$\tau_{ep}$}
\newcommand{\tauee}{$\tau_{ee}$ }
\newcommand{\taueey}{$\tau_{ee}$}
\newcommand{\jphi}{$j_{\phi}$ }
\newcommand{\jphiy}{$j_{\phi}$}
\newcommand{\tc}{$T_{c}$ }
\newcommand{\tcy}{$T_{c}$}
\newcommand{\hcl}{$H_{c1}$ }
\newcommand{\hcly}{$H_{c1}$}
\newcommand{\ef}{$E_{F}$ }
\newcommand{\efy}{$E_{F}$}
\newcommand{\estar}{$E^{*}$ }
\newcommand{\estary}{$E^{*}$}
\newcommand{\htc}{high-temperature superconductors } 
\newcommand{\htcy}{high-temperature superconductors}
\newcommand{\et}{{\it et al. }}
\newcommand{\ety}{{\it et al.}}
\newcommand{\be}{\begin{equation} }
\newcommand{\ene}{\end{equation}}
\newcommand{\hh}{$H$ }
\newcommand{\hhy}{$H$}
\newcommand{\hc}{$H_{c}$ }
\newcommand{\hcy}{$H_{c}$}
\newcommand{\ho}{$H_{0}$ }
\newcommand{\hoy}{$H_{0}$}
\newcommand{\jc}{$j_{c}$ }
\newcommand{\jcy}{$j_{c}$}
\newcommand{\sg}{superconducting }
\newcommand{\sgy}{superconducting}
\newcommand{\ssc}{superconductor }
\newcommand{\sscy}{superconductor}
\newcommand{\hcu}{$H_{c2}$ }
\newcommand{\hcuy}{$H_{c2}$}
\newcommand{\rfff}{$\rho_{f}$ }
\newcommand{\rfffy}{$\rho_{f}$}
\newcommand{\hcut}{$H_{c2}(T)$ }
\newcommand{\hcuty}{$H_{c2}(T)$}
\newcommand{\jd}{$j_{d}$ }
\newcommand{\jdy}{$j_{d}$}
\newcommand{\id}{$I_{d}$ }
\newcommand{\idy}{$I_{d}$}
\newcommand{\ybco}{Y$_{1}$Ba$_{2}$Cu$_{3}$O$_{7}$ }
\newcommand{\ybcoy}{Y$_{1}$Ba$_{2}$Cu$_{3}$O$_{7}$}
\newcommand{\lsco}{La$_{2-x}$Sr$_{x}$CuO$_{4}$ }
\newcommand{\lscoy}{La$_{2-x}$Sr$_{x}$CuO$_{4}$}
\newcommand{\mgb}{MgB$_{2}$ }
\newcommand{\mgby}{MgB$_{2}$}
\newcommand{\de}{$\delta \epsilon$ }
\newcommand{\dey}{$\delta \epsilon$}
\newcommand{\nq}{$n_{q}$ }
\newcommand{\nqy}{$n_{q}$}
\newcommand{\rrhon}{$\rho_{n}$ }
\newcommand{\rrhony}{$\rho_{n}$}
\newcommand{\rrho}{{$\rho$} }
\newcommand{\rrhoy}{{$\rho$}}
\newcommand{\qp}{quasiparticle }
\newcommand{\qpy}{quasiparticle}
\newcommand{\qps}{quasiparticles }
\newcommand{\qpsy}{quasiparticles}
\newcommand{\bib}{\bibitem}
\newcommand{\ib}{{\em ibid. }}
\newcommand{\taue}{$\tau_{\epsilon}$ }
\newcommand{\tauey}{$\tau_{\epsilon}$}
\newcommand{\vstary}{$v^{*}$}
\newcommand{\vstar}{$v^{*}$ }
\newcommand{\tstar}{$T^{*}$ }
\newcommand{\tstary}{$T^{*}$}
\newcommand{\rhostar}{$\rho^{*}$ }
\newcommand{\rhostary}{$\rho^{*}$}
\newcommand{\vinf}{$v_{\infty}$ }
\newcommand{\vinfy}{$v_{\infty}$}
\newcommand{\fd}{$F_{d}$ }
\newcommand{\fdy}{$F_{d}$}
\newcommand{\fe}{$F_{e}$ }
\newcommand{\fey}{$F_{e}$}
\newcommand{\fl}{$F_{L}$ }
\newcommand{\fly}{$F_{L}$}
\newcommand{\jstar}{$j^{*}$ }
\newcommand{\jstary}{$j^{*}$}
\newcommand{\je}{$j(E)$ }
\newcommand{\jey}{$j(E)$}
\newcommand{\vphi}{$v_{\phi}$ }
\newcommand{\vphiy}{$v_{\phi}$}
\newcommand{\blo}{$B_{1}$ }
\newcommand{\bloy}{$B_{1}$}
\newcommand{\bhi}{$B_{\infty}$ }
\newcommand{\bhiy}{$B_{\infty}$}
\newcommand{\vlo}{$v_{1}$ }
\newcommand{\vloy}{$v_{1}$}
\newcommand{\bo}{$B_{o}$ }
\newcommand{\boy}{$B_{o}$}
\newcommand{\eo}{$E_{o}$ }
\newcommand{\eoy}{$E_{o}$}

\preprint{Submitted to Phys. Rev. Lett.}

\title{The pair-breaking critical current density of magnesium diboride}

\author{Milind N. Kunchur} 
 \homepage{http://www.physics.sc.edu/kunchur}
 \email{kunchur@sc.edu}
\affiliation{Department of Physics and Astronomy\\
University of South Carolina, Columbia, SC 29208}

\author{Sung-Ik Lee}
\author{W. N. Kang}
\affiliation{Department of Physics, Pohang University of Science and
Technology\\Pohang 790-784, Republic of Korea}

\date{\today}

\begin{abstract}
The pair-breaking critical current density, $j_{d}$, of magnesium
diboride was determined over its entire temperature range by a pulsed dc
transport measurement. At fixed low values of current density $j$, the 
resistive transition temperature \tc 
shifts in the classic $\Delta T_{C}(j)/T_{C}(0) \propto
-[j/j_{d}(0)]^{2/3}$ manner, with a projected $j_{d}(0) \approx 2
\times 10^{7}$ A/cm$^{2}$. The directly measured $j_{d}(0)$, from 
current-voltage ($I$-$V$)
curves at different fixed temperatures, has a similar value 
and the overall temperature dependence  $j_{d}(T)$ 
and magnitude are consistent with Ginzburg-Landau theory. 

\end{abstract}

\maketitle

\section{\label{sec:level1}Introduction}

Magnesium diboride (\mgby ) recently made an impact as 
a promising new superconductor with a surprisingly 
high critical temperature 
for a simple binary compound. This has spurred considerable 
research activity into investigating the myriad properties associated
with its superconducting state. Besides the critical temperature \tc and
the upper critical field \hcuy , an intrinsic parameter that 
sets a fundamental limit to the survival of superconductivity is the
pair-breaking (or depairing) critical current density \jdy . 
We report the first measurement of this important quantity in the \mgb
superconductor, which sets an absolute limit to the maximum
current-carrying performance under ideal conditions. This also
represents, to our knowledge, the only complete ($0 \alt T \alt$ \tcy )
measurement of \jd by a direct transport method in any type-II
superconductor.

When a superconducting state is formed, charge carriers 
correlate and condense into a coherent macroscopic quantum state. 
The formation of this state  
is governed principally by a competition between four energies: condensation,
magnetic-field expulsion, thermal, and kinetic. The order
parameter $\Delta$, that
describes the extent of condensation and the strength of the 
superconducting state, is
reduced as the temperature $T$, magntetic field $H$, and electric current
density $j$ are increased.
The boundary in the $T$-$H$-$j$ phase space that separates the 
superconducting and normal states
is where $\Delta$ vanishes, and the three parameters attain their
critical values $T_{c}(H,j)$,
$H_{c2}(T,j)$, and $j_{d}(T,H)$. 

In practice, a superconductor loses its ability to carry resistanceless 
current long before $j$ reaches \jdy . Any process that causes the
phase difference between two points to change with time---such as the motion
of flux vortices, phase slip centers in narrow wires, junctions, and
fluctuations---can generate a finite voltage and hence resistance. 
The conventional critical-current density \jc marks this onset of
dissipation---depending on extrinsic variables such as 
vortex pinning by defects---and in type-II superconductors can be
a few orders of magnitude lower than \jdy . Thus the transport becomes
resistive and intensely dissipative long before the thermodynamic limit 
is reached, tending to mask a direct measurement of \jd by
sample heating. In this
work we use a highly evolved pulsed-current technique (which we have refined
over ten years) to reduce heating and obtain 
\jdy$(T\rightarrow 0)$ both from (1) a direct measurement of $j$ required
to drive the system normal at $T \ll $\tc and (2) from the shift in \tc
as a function of $j$ near $T \sim $\tcy . 
\section{Theoretical background} 
A theoretical estimate of \jd can be obtained from the 
Ginzburg-Landau (GL) theory, in which the strength of the
superconducting state is expressed through the complex phenomenogical
order parameter $\psi=|\psi|e^{i \varphi}$. 
The superfluid density near \tc is proportional
to $|\psi|^{2}$ and the free-energy density $f$ of the system 
(w.r.t. the free-energy density in the normal state) 
can be expressed as a power expansion in  $|\psi|^{2}$
(In ``dirty'' superconductors---superconductors with a high impurity
scattering rate---the approximate validity of the GL expressions extends down
to $T \ll$ \tcy .). 
In the absence of significant magnetic fields and in situations where
the magnitude of the order parameter  $|\psi|$ is uniform (either because
the dimensions of the sample are small compared to the coherence length
or because of the principle of minimum entropy production at high
dissipation levels\cite{metal}) 
$f$ can be expressed as\cite{tinkhamtext}
\begin{equation} \label{free}
f=
\alpha |\psi|^{2} + \frac{\beta}{2} |\psi|^{4} +
\frac{1}{2}|\psi|^2 m^{*}v_{s}^{2}. \end{equation}
$\alpha$ and $\beta$ are negative and positive constants respectively 
($\alpha$ becomes positive above \tcy ), and 
the positive third term 
is the kinetic energy density 
expressed in terms of the superfluid velocity
$\mbox{\boldmath $v$}_{s}=\frac{\hbar \nabla \varphi}{m^{*}}
-\frac{e^{*}\mbox{\boldmath $A$}}{cm^{*}}$; where $e^{*}$ 
and $m^{*}$ are respectively the
effective charge and mass of a cooper pair. 
For zero $v_{s}$, the equilibrium value of $|\psi|^{2}$ that minimizes the free
energy (Eq.~\ref{free}) is $|\psi_{\infty}|^{2}=-\alpha/\beta$. 
For a finite $v_{s}$ it becomes 
\begin{equation}
|\psi|^{2}=|\psi_{\infty}|^{2} \left( 1 - \frac{m^{*}v_{s}^{2}}{2
|\alpha|} \right). \end{equation}
The corresponding supercurrent density is 
\begin{equation} 
j= e^{*}|\psi|^{2}
v_{s}
= 2e |\psi_{\infty}|^{2}
\left(1 -
\frac{m^{*}v_{s}^{2}}{2 |\alpha|} \right) v_{s}. 
\end{equation}
The maximum possible value of this expression can now be identified with
$j_{d}$: 
\begin{equation} j_{d}(T)= 2e |\psi_{\infty}|^{2}
\frac{2}{3}\left(\frac{2\alpha}{3m^{*}} \right)^{1/2} = 
\frac{c H_{c}(T)}{3\sqrt{6} \pi \lambda (T)}
\end{equation}
where the GL-theory parameters were replaced by their 
expressions $\alpha(T)=-(e^{*2}/m^{*}c^{2})H_{c}^{2}(T)\lambda^{2}(T)$ and 
$\beta(T)=(4\pi e^{*4}/m^{*2}c^{4})H_{c}^{2}(T)\lambda^{4}(T)$ in terms
of the physically measureable quantities 
$H_{c}$ (thermodynamic critical
field) and $\lambda$ (magnetic penetration depth). 
The relations $H_{c}(T) \approx H_{c}(0)[1-(T/T_{c})^{2}]$ and 
$\lambda(T) \approx \lambda(0)/\sqrt{[1-(T/T_{c})^{4}]}$ give
\begin{equation} \label{jdtfull} 
 j_{d}(T) \approx j_{d}(0) [1-(T/T_{c})^{2}]^{\frac{3}{2}} 
[1+(T/T_{c})^{2}]^{\frac{1}{2}} 
\end{equation}
where 
\begin{equation} \label{jdtzero}
j_{d}(0)=cH_{c}(0)/[3\sqrt{6}\pi \lambda(0)] \end{equation}
is the zero-temperature depairing current density. 
Close to \tcy , Eq.~\ref{jdtfull} reduces to 
$ j_{d} (T \approx T_{c}) \approx 
4j_{d}(0)[1-T/T_{c}]^{\frac{3}{2}}$. This can be inverted to give 
the shift in transition temperature
$T_{c}(j)$ at small currents, with the well-known $j^{2/3}$ proportionality:
\begin{equation} \label{tcjsmall}
\frac{T_{c}(0)-T_{c}(j)}{T_{c}(0)} \approx 
\left(\frac{1}{4}\right)^{\frac{2}{3}} 
\left[\frac{j}{j_{d}(0)}\right] ^{\frac{2}{3}}. \end{equation}
(The  preceding discussion is based on Refs. 
\cite{tinkhamtext} and \cite{bardeen}.) 
Note that if heat removal from the sample is ineffective, 
Joule heating will give an apparent shift $\Delta T_{c} \propto \rho
j^{2}$, which is the cube 
of the intrinsic $\sim j^{2/3}$ depairing shift near \tcy , and hence easily
distinguishable. 

\section{Experimental details}
The samples are 400 nm thick films of MgB$_{2}$ 
fabricated using a two-step method whose details are described 
elsewhere\cite{sampleprep,sampleprep2}. An amorphous boron film was 
deposited on a (1\={1}02) Al$_{2}$O$_{3}$ substrate at room temperature by 
pulsed-laser ablation. The boron film
was then put into a Nb tube with high-purity Mg metal (99.9\%)
and the Nb tube was then sealed using an arc furnace in an
argon atmosphere. Finally, the heat treatment was carried out at
900$^{\circ}$ C for 30 min. in an evacuated quartz ampoule 
sealed under high vacuum. X-ray diffraction indicates 
a highly c-axis-oriented crystal structure normal to 
the substrate surface with no impurity phases.
The films were photolithographically patterned down to narrow bridges.
In this paper we show data on three bridges, labelled S, M, and L (for
small, medium, and large) with lateral dimensions 
2.8 x 33, 3.0 x 61, and 9.7
x 172 $\mu$m$^{2}$ respectively. The lateral
dimensions are uncertain by $\pm 0.7 \mu$m and the thickness by $\pm 50$ nm.

The electrical transport measurements were made using a pulsed signal source
with pulse durations ranging 0.1--4 $\mu$s and a duty cycle of about 1 ppm. 
From past experience with other films (e.g., \ybco on LaAlO$_{3}$)
we found that micron-wide 
bridges typically have thermal resistances
of order $R_{th} \sim$ 1--10 nK.cm$^{3}$/W at
microsecond timescales \cite{unstable,metal,mplb}.  For the
present film-substrate combination, complete information about the thermal
constants was not available to calculate \rth from first principles, but
we are able to show by other means that sample heating is not appreciable.
Further details of the
measurement techniques have been published in a previous review article
\cite{mplb} and other recent papers \cite{metal,unstable}. All
measurements were made in zero applied magnetic field and the
highest self field of the current ($\sim 300$ G) is of the order of  
the lower-critical field \hcly =185 G \cite{wang}.

\section{Results and analysis} 
\begin{figure}[h] 
\begin{center}
	\leavevmode
	\epsfxsize=\hsize
	\epsfbox{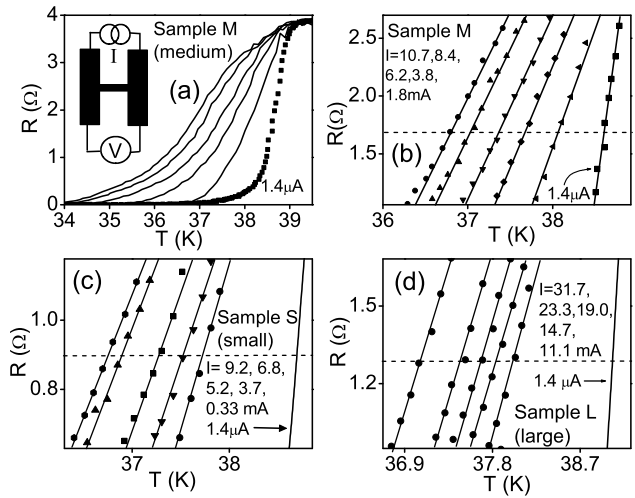}
\end{center}
\vspace{-1em}
\caption{\label{rtcurves}{\em Resistive transitions of \mgb bridges at
different currents (values correspond to curves from left to right.). 
Panels (a) and (b) show two windows of the same data. The inset in (a) shows
the sample geometry and configuration of 
leads. Panels (b), (c), and (d) show the
central main portions of the transitions for three different sized 
samples. The rightmost curves at I=1.4 $\mu$A were measured by with a
continuous DC current; the rest used 
pulsed signals. The dashed lines represents $R=R_{n}/2$ for each sample.}}
\end{figure}

Fig.~\ref{rtcurves}(a) shows the resistive transitions at different electric
currents $I$
for the medium sample. The inset shows the sample
geometry. The horizontal sections of the current leads add a small
($\sim 15$ \%) series resistance 
to the actual resistance of the bridge. Because $j$ in these
wide regions is negligible, this resistance freezes out at the nominal
unshifted \tcy ,  making the onset seem to not shift. However over the
main portions of the curves, there are substantial and relatively
parallel shifts induced by current. Fig.~\ref{rtcurves}(b) provides a
magnified view of the central two-thirds portion of the transitions. The
dashed line represents half the normal-state resistance ($R_{n}$) of the
bridge, which serves as the criterion (resistive-transition-midpoint)
for defining \tcy $(j)$.
Panels (c) and (d) show similar sets of
curves for the other two (small and large) samples. 
\begin{figure}[h] 
\begin{center}
	\leavevmode
	\epsfxsize=0.7\hsize
	\epsfbox{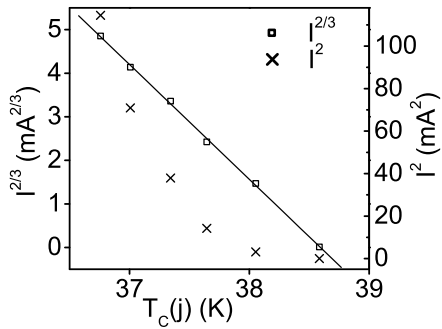}
\end{center}
\vspace{-1em}
\caption{\label{laws}{\em 
Shifted transition temperatures 
at different currents. The two y-axes plot the
same \tcy $(j)$ data versus $I^{2/3}$ and  $I^{2}$, showing adherance to
the  $I^{2/3}$ law for pair breaking rather than the
$I^{2}$ law for Joule heating. The linear fit (solid line) to the  
$I^{2/3}$ plot gives $I_{d}(0)= 257$ mA (see Eq.~\ref{tcjsmall}).}}
\end{figure}

Fig.~\ref{laws} shows the midpoint \tcy 's 
and their corresponding currents (ranging 
from $10^{-6}$ to $10^{-2}$ A) plotted as $I^{2/3}$ (expected for
pair-breaking) and as $I^{2}$ (expected for Joule heating).
 The shifts are closely proportional to $I^{2/3}$ 
rather than to $I^{2}$, showing that heating is not appreciable 
(the plots for samples S and L look similar).  
The slope $d I^{2/3}/dT_{c}(j)$ together with Eq.~\ref{tcjsmall} gives
a zero-temperature depairing current value of 257 mA. Dividing this by the
cross-sectional area gives a current density 
of \jdy $(0)=2.1 \times 10^{7}$ A/\cmsy . 
The respective values for samples S and L are \jdy $(0) = 
2.2 \times 10^{7}$ and $1.8 \times 10^{7}$ A/\cmsy . 
The three values are consistent within the uncertainities in 
the sample dimensions, implying a cross-sectionally uniform current density.
This is expected 
for the dissipative state of a superconductor (In the fluctuation
region near the \tcy$(j)$ boundary and during
flux motion--when the superconductor is resistive---the 
current flow becomes macroscopically uniform, as in a normal conductor, due
to the principle of minimum entropy production. This has been discussed
and verifed elsewhere \cite{metal}.) and close to the \tcy$(j)$ boundary
where $\lambda$ and $\xi$ (coherence length) diverge. 

\begin{figure}[h] 
\begin{center}
	\leavevmode
\begin{tabular}{cc}
	\epsfxsize=0.52\hsize 
	\epsfbox{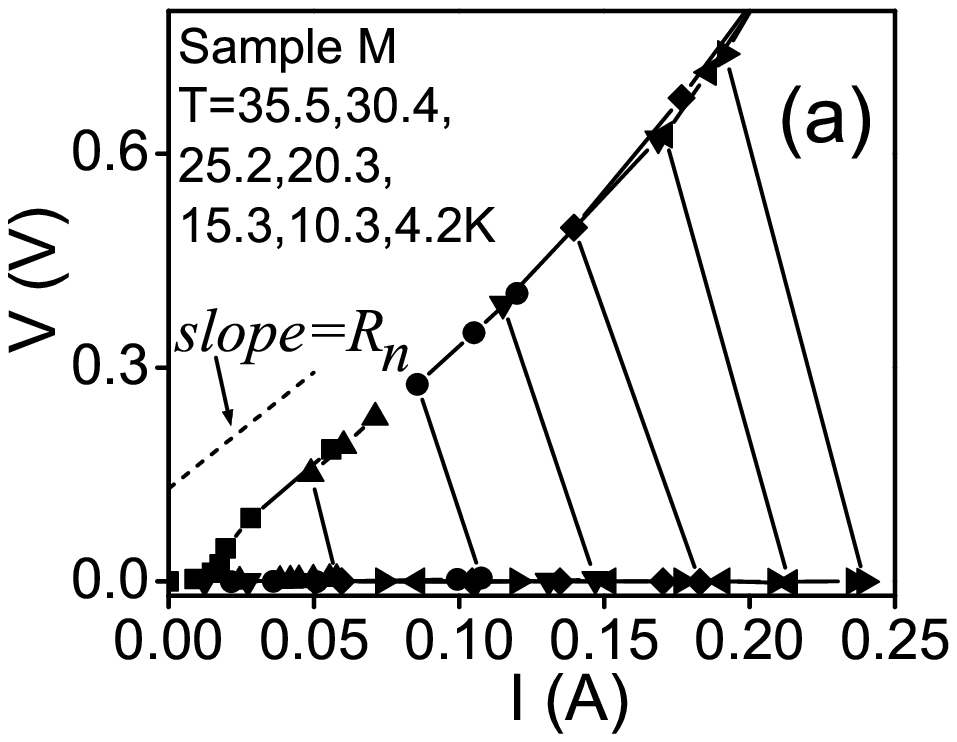} \hspace{-2em}&
	\epsfxsize=0.51\hsize \hspace{-2em}
	\epsfbox{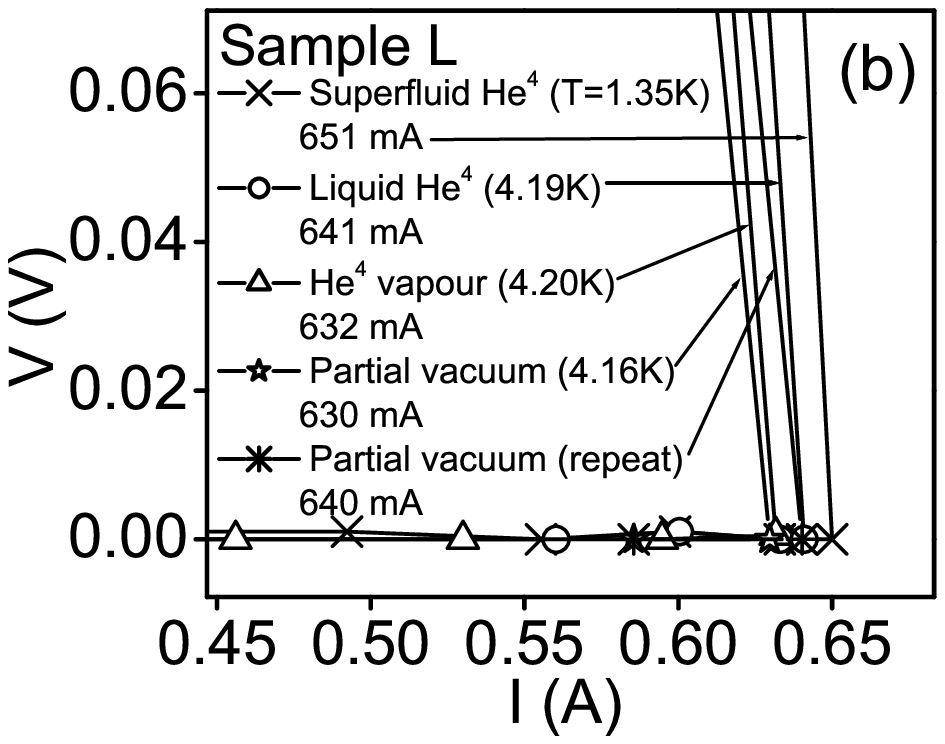}
\end{tabular}
\end{center}
\vspace{-1em}
\caption{{\em (a)
$IV$ curves for sample M at seven fixed temperatures (listed for curves going from left to
right). Beyond \jdy , the voltage jumps to a
linear behavior reflecting the resistance of the normal state 
(slope indicated by the dashed line). (b) $IV$ curves for the largest
sample in different thermal environments to evaluate Joule heating.}}
\label{ivcurves}
\end{figure}
Fig.~\ref{ivcurves}(a) shows current-voltage ($IV$) characteristics at various
fixed temperatures for sample M (results for samples S and L are similar).
 As $I$ is increased $V$ remains close to zero until some
critical value. Above this it shows Ohmic behavior $V=I R_{n}$.
Note that at $T$=35.5K the transition is
gradual, whereas at the lower temperatures it is rather abrupt. This is
in part because a type II superconducting phase transition changes from
second order to first order at lower temperatures in the presence of a
current \cite{bardeen} and possibly because of a thermal component. The
``s'' shape arises because the external circuit feeding the pulsed signal has 
a source impedance $R_{s}$ of about 12 $\Omega$. Thus
when the sample is driven normal, the current will drop discontinuously by
the fraction $R_{n}/(R_{n}+R_{s}) \sim 20\%$ as observed. Although the
amount of Joule heating cannot be directly estimated, its significance can be
assessed by measuring the $IV$ curves in different thermal
environments. The previous curves in Fig.~\ref{ivcurves}(a) were all
measured with the sample in helium vapour. Such measurements were
repeated with the samples in superfluid and normal liquid helium, and in
vacuum, and Fig.~\ref{ivcurves}(b) shows one such set for the large
sample (its lower surface-to-volume ratio gives it the worst
thermal resistance). 
Fig.~\ref{ivcurves}(b) shows no significant systematic influence
of the environment on the observed value of \jdy , which would not be
the case if Joule heating were a serious problem \cite{huebener}. 

From such $IV$ characteristics measured at the lowest temperature (1.5 
K) in superfluid helium, the current required to drive the sample normal
provides a direct lower bound on \jdy$(T\approx
0)$ (This lower bound will equal \jd in the case of uniform current flow.
We return to this point again later.). 
 For the three samples S, M, and L these respective values are 
\jdy$(0) = 1.9 \times 10^{7}$, $2.0 \times 10^{7}$,
 and  $1.7 \times 10^{7}$ A/\cmsy , which are consistent with the values
obtained earlier ($2.2 \times 10^{7}$, $2.1 \times 10^{7}$, 
and $1.8 \times 10^{7}$ A/\cmsy ) from the shifts in the resistive
transitions near \tc (Fig.~\ref{laws} and Eq.~\ref{tcjsmall}).  
\begin{figure}[h] 
\begin{center}
	\leavevmode
	\epsfxsize=0.7\hsize
	\epsfbox{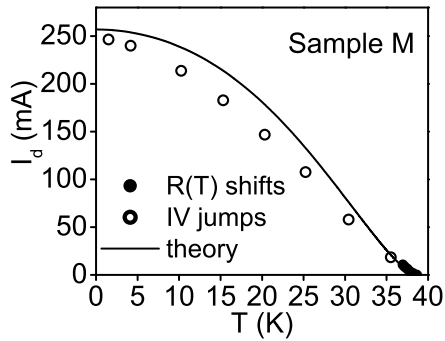}
\end{center}
\vspace{-1em}
\caption{{\em Pair-breaking currents from IV curves (Fig.~\ref{ivcurves})
and \tc shifts (Figs.~\ref{rtcurves} and \ref{laws}). 
The solid line represents the theoretical
curve (Eq.~\ref{jdtfull}) in which $I_{d}(0)$ and \tc are fixed by the
\tc shifts (Fig.~\ref{laws}; $36 < T < 39$), with no
adjustment made over the rest of the range ($1.5 \leq T \leq 36$ K).}}
\label{glcurve}
\end{figure}

Fig.~\ref{glcurve} 
shows the values of \id at different temperatures obtained from the $IV$ 
characteristics of Fig.~\ref{ivcurves}(a). Also shown are the
values of \id obtained from the shifts in the resistive transition near
\tc (from Fig.~\ref{laws}). The solid line
is a plot of Eq.~\ref{jdtfull} in which the values of \tcy$(0)$ and
\idy$(0)$ came directly from the observed $j^{2/3}$ behavior 
of Fig.~\ref{laws} and were not adjusted to fit the other data over the
extended temperature range, nor was \idy$(0)$ adjusted to fit the actual
measured value from the $IV$ characteristic at low $T$ (i.e.,
Fig.~\ref{ivcurves}). Nevertheless, the \idy $(T)$ data tend to 
follow the general trend of Eq.~\ref{jdtfull}.

An aspect of these data that may seem surprising is that
even for $T \ll$ \tc (where the sample width 
$w \gg \lambda, \xi$), the average 
current density in the bridge reaches essentially the full \jd before
the system becomes normal. 
By contrast, previous studies \cite{skocpol} 
of \jd in wide low-\tc type-I bridges found that the current
distribution was non-uniform and the sample was driven normal 
when the peak $j$ near the surface exceeded \jdy . Using their model 
with our sample dimensions and parameters, our effective \jd should have been 
reduced by a factor of 3, but it is not. In our present
relatively high-\tc type II material, it seems that the slight flux
motion induced by the self field and fluctuations (because of the
proximity to the \tcy $(j)$ phase boundary and much higher $T$) 
serve to homogenize the current distribution. Ironically, such incipient 
dissipation may actually stabilize the flow and permit the average $j$ to get
closer to \jd before the system becomes normal. 

In conclusion, we have measured the fundamental
pair-breaking critical current density of magnesium diboride over the
entire temperature range for in-plane current transport. 
The measured \jdy $(T)$ function is 
consistent with the Ginzburg-Landau form and conforms exactly 
to the $\Delta T_{c} \propto j^{2/3}$ behavior predicted near \tcy . 
\jdy $(0)$ obtained from the value of current required to drive the 
sample normal at  $T \rightarrow 0$, agrees with the \jdy $(0)$ deduced from
the  $\Delta T_{c} \propto j^{2/3}$ behavior close to \tcy .
The average value for all samples by both methods is \jdy $(0) \approx 
1.9 \pm 0.4 \times 10^{7}$ A/\cmsy . This is comparable in order of
magnitude to the value of $6.1 \times 10^{7}$ A/\cms calculated from 
Eq.~\ref{jdtzero} and the 
published values of 
\hcy =2500 G and $\lambda = 185$ nm from the review on
\mgb by Wang et al. \cite{wang}, in
view of the uncertainities in those parameters. From a technological
standpoint, the depairing current density of \mgb is about an order
of magnitude lower than the high-\tc cuprates \cite{pair}. The good news
is that the flux pinning in films is so strong (because of the larger coherence
length and more isotropically 3-D behavior) that the depinning \jc 
at modest fields appear to be within an order of magnitude of
\jd \cite{otherhighjs}, whereas for the cuprates \jc and \jd can be 
separated by two or three orders of magnitude\cite{mgbfluxflow}.

\section{Acknowledgements}
The authors acknowledge useful discussions and other assistance from
J. M. Knight, B. I. Ivlev, C. Wu, D. H. Arcos, 
H.H.Acros, H.J. Kim, E.M. Choi, K.J. Kim, and D. K. Finnemore. 
This work was supported by the U. S. Department of Energy 
through grant number DE-FG02-99ER45763 and by
the Creative Research Initiatives of the Korean Ministry of Science
and Technology.




\begin{references}
\bibitem{mgbfluxflow} Preliminary in-field measurements on \mgb 
at low temperatures indicate $IV$
curves that are highly non-linear with no Ohmic free-flux-flow behavior at
all. Instead the resistivity rises rapidly from its onset to its full 
normal-state value due to a combination of pair 
breaking and flux motion. Details of this non-linear flux flow
behavior will be discussed elsewhere.
\end{references}
\end{document}